# The fluid motion physics:
## The interaction mechanics of a free liquid jet with a body and with the other free liquid jet

S.L. Arsenjev[1]

*Physical-Technical Group*
*Dobroljubova Street, 2, 29, Pavlograd Town, Dniepropetrovsk region, 51400 Ukraine*

Solution of a problem on the interaction mechanics of a free liquid jet with a flat plate, body and with other jet has been achieved by means of a graphic-analytical method, developed by author of the given article. This method has allowed physically adequately and visually to describe the flow field near the streamlined surface and to give expressions for quantitative evaluation of the jet pressure profile onto this surface. This method is equally correct for both a flat jet and a jet with a round cross-section. Analysis of the flow field has allowed to detect a jet component, induced by the body fore part in the unrestricted fluid stream and determining the body form drag. Besides that, it has been ascertained that a friction also induces the jet component in the potential boundary layer. It has been introduced a new notion of the stream total head vector, determining an origin of the possible jet flow in the stream.
**PACS:** 01.55.+ b; 05.65.+ b; 47.10.+ g; 47.27.Wg; 47.85.Dh.

## Introduction

A subject of research in the given article is a problem on the interaction mechanics of the compact free liquid jet – for example the water jet in atmosphere – with the flat plate.
A history of this question: the most early publication on the results of experimental research in the frameworks of the mentioned problem is contained in a book [1], written by P.L.G. Du Buat (1786) who was the talented French experimenter and theorist in a field of hydromechanics. Du Buat, side by side with J.C. Borda, is one of those who had derived the correct expression for the outflow equation $v = \sqrt{2gh}$. In his own experimental researches Du Buat for the first time had detected that the resistance force in movement of a flat plate, submerged into water, is sufficiently greater than the pressure force of the free water jet onto that plate under the same relative velocity (so-called Du Buat's paradox).
The state-of-the-art of the problem: a carrying out of the not complicated experiments and the simplest measurements allow to determine both the pressure force quantity of a water flat jet onto the flat plate under various angles of the plate inclination to the jet axis and the weight (volume) flow components at a spreading of the jet into two streams. An analysis of these experiments allows quantitatively to express its results by means of two equations: the momentum conservation and the continuity condition. The problem well-known solution is restricted by consideration of a flat jet only. This solution does not give the coordinate of the jet pressure centre on the plate and one does not give explanation why the jet pressure force onto the plate is proportional to the jet doubled velocity head. The overcoming of the listed shortcomings by means of creation of the phenomenon conceptual model and then of its mathematical description is a basis of the given article contents.

## Approach

Author of the given article assumes the Physical Ensemble method, stated in articles [2, 3], written by him self-dependently, and in articles [4 - 12], written by him jointly with his collaborators of the Physical-Technical Group, as a basis for the given problem solution.

[1] Phone: (+38 05632) 40596 (home, Rus.)
E-mail: ptglecus@yahoo.com

Conformably to the problem the mentioned method envisages:
- elucidation of geometrical structure of the free liquid jet - in the simplest case of a flat jet - in its interaction with the flat plate - in the simplest case oriented perpendicularly to the jet axis;
- an use of the well-known results of experimental determination of the jet pressure force onto the plate in its quantity, direction of its action and a point its application;
- creation of the conceptual model of the interaction and then creation on this basis of its mathematical description, allowing to consider the diversity of the interaction variants of the free liquid jet with the flat plate, as well as of such jets between themselves and the other cases.
A development of graphic-analytical method for solution of the stated problem is stipulated by some circumstances. In the first, a solving power of combination of geometrical constructions with its analytical description surpasses a solving power of each of these methods, taken separately. In the second, the basic formulas of mechanics, describing the uniformly accelerated (decelerated) motion, for example a free fall,

$$v = gt; \qquad v = \sqrt{2gh}; \qquad h = g\,t^2/2,$$

are also true for hydromechanics and ones have united geometrical interpretation in the kind of the following accordance: $v$ is an arc length of the circumference by radius $g$ at central angle $t$; $h$ is an arc length of an evolvent of the above mentioned circumference as of its evolute.
Fig.1 shows the geometrical construction, containing the first quarter of the evolutic circumference and its evolvent. The arc length of this evolvent is placed along the axis of abscissa and the corresponding curvature radii of this evolvent are placed along the axis of ordinates in the same place. A line, passing through the ends of the evolvent radii, is the evolventic parabola. Focal parameter of this parabola – a distance from its focus up to its directrix – is equal to the evolutic circumference radius. One cannot but admit that the evolvent of the first quarter of the circumference is at the same time a quarter of an oval quite like an ellipse with a ratio of its axes equal $\pi/2$ (or $2/\pi$). The arc length of this evolventic oval quarter is determined by a simple expression $h = g(\pi/2)^2/2$ in contrast to the rest of the ellipse modifications whose length can be determined by means of the elliptic integral only. Geometrical affinity of the evolutic circumference, the evolventic parabola and the evolventic oval is that these curved lines are the so-called conical sections. At the same time the evolutic circumference is also trigonometric circle as a basis for a constructing of graph of sinusoidal quantities.
Combination of these elementary curves is assumed as a basis of the graphic-analytical method for solution of problems on interaction of the free liquid jets with bodies, of these jets between themselves and with the unrestricted stream of homogeneous liquid.

**Solution**

According to the first thesis of the above stated approach an observation of the experiment allows to ascertain the following:
- the initial straight flat jet spreads mirror-symmetrically into two equal streams, moving along the plate surface in opposite directions; a thickness of the streams is the same and equal to a half of the initial jet thickness;
- a division of initial jet into two equal streams is going on smoothly so that the jet outside profile is similar to a circumference arch; radius of this arch is equal to a half of the initial jet thickness.
According to the second thesis of the above stated approach the well-known results of experimental research testify to that the jet pressure force acts perpendicularly to the plate surface and its quantity is equal to the doubled product of the jet velocity head and the jet cross-section area before zone of its interaction with the plate.
Now it is necessary to elucidate the question why the jet pressure force on the plate is twice as much of the jet force, stipulated by its velocity head. To resolve this question it is necessary to appeal to the experimental results. Equality of a turn radius of the jet external surface to a half thickness of initial jet in zone of its immediate interaction with the plate means that a breadth of the jet turn zone on the plate is equal to a doubled thickness of the initial jet.

Fig.2*a* shows a diagram of a spreading of the free liquid jet (with parameters: $\rho$ is the liquid mass density; $v_0$ is the jet motion velocity; $2b$ is the jet thickness) on the flat plate *AB*, oriented perpendicularly to the jet axis. A profile of the jet external surface is mirror-symmetrically turned in both sides from its axis with radius equal to a half the jet initial thickness. It can be supposed, that the jet internal part spreads equidistantly to a profile of the jet external surface in the kind of the circumference arches *CD* and *CE*. Suppose also, that the liquid motion in zone of a triangle *CDE* (with two compasses-curvilinear sides) is sufficiently braked, therefore the all weight (volume) flow of the jet is realized in its two compasses-curvilinear branches, containing between a profile of the jet external surface and lines *CD* and *CE* equidistant to it.

Thus, if a static pressure, stipulated by the jet velocity head and equal to it in a quantity, acts in the triangle *CDE*, then the pressure must be distributed on the line *CE* with length a twice as much than the initial jet thickness.

Now it is necessary to elucidate a question on possibility of a supporting of the mentioned static pressure on a level of the jet velocity head in the triangle *CDE* zone. To resolve this question it is necessary to pay attention to the circulative character of a motion of the jet branches from a section $O_1O_2$ up to sections $O_1E$ and $O_2D$, as well as to take into account an equality of the motion linear velocity on the internal and external boundaries of the jet branches and equality of its velocity to the initial jet velocity. With the taking into account of these features the well-known expression for determination of a pressure, stipulated by rotation of a liquid together with a vessel,

$$p_c = \rho \frac{\omega^2}{2}\left(r^2 - r_0^2\right)$$

takes the shape

$$p_c = \rho \frac{\omega^2}{2}\left(r^2 - r_0^2\right) + \rho \frac{\omega^2}{2}\cdot r_0^2, \qquad (1)$$

hence with the considering that $\omega = v_0 / r$ it follows

$$p_c = \rho \frac{v_0^2}{2}.$$

Thus, the centrifugal forces, acting in the jet branches, ensure an equality of the static pressure in the triangle *CDE* zone to the initial jet velocity head. As a result, an action of this static pressure onto the plate by breadth *DE*, which one is a twice as much then the jet thickness, ensures the jet pressure force onto the plate a twice as much then a product of the initial jet velocity head and its cross-section area.

Figs.2*b, c* show diagrams of a spreading of the flat free liquid jet on the flat plate *AB*, oriented to the jet axis under angles $60^0$ and $30^0$ accordingly. The well-known results of experimental researches testify to that the jet pressure force onto the plate, oriented under angles within the limits $90^0 > \alpha > 0$, is changed in proportion to $\sin\alpha$ ; at the same time this pressure force is displaced aside from the jet axis as it is shown in fig.2*b, c*. With the taking into account of the above stated results of the well-known experimental research and the analysis of these results the method of a constructing of the diagram of the jet spreading is the following:

- direct the longitudinal profile lines of the flat free liquid jet of the given thickness and under given angle to the flat plate up to a crossing with the plate surface track *AB*;
- construct a quadrant on the initial jet axis and its longitudinal profile;

- drop a perpendicular out of the quadrant centre on the plate surface track $AB$;
- draw a straight line parallel to the jet axis from the quadrant arch point up to the plate surface track $AB$;
- draw a semi-circumference with a radius equal to the initial jet thickness and with a centre in the intersection point of the straight line parallel to the jet axis with the plate surface track; marc off the point $C$;
- rise two perpendiculars to the plate surface track out of the semi-circumference diameter (points $E$ and $D$);
- draw a straight line perpendicularly to the initial jet axis through the point $C$ up to intersection with two perpendiculars out of the points $E, D$; mark off the points $O_1$ and $O_2$;
- use the point $O_1$ and $O_2$ as the centers of arches of circumferences for a constructing of contours of the jet branches.

The figure $O_1O_2DE$ restricts a zone of the immediate division of the initial jet into its two branches. The curvilinear triangle $CDE$ restricts a zone where the jet flow is sufficiently braked. Quantity of the static pressure in the triangle $CDE$ zone depends on the plate orientation angle relatively the initial jet motion direction in accordance with expression

$$p_c = \rho \, \frac{v_0^2}{2} \, \sin\alpha \, . \qquad (2)$$

At the same time inequality of thickness of the jet branches and its curvature radii give rise to doubt concerning possibility of a supporting of the static pressure of the same quantity in the triangle $CDE$ zone. To dispel this doubt it is necessary to consider an action of the centrifugal forces every branch of the jet. For simplicity it can be used an expression for the centrifugal force, accepted in theoretical mechanics

or

$$F_1 = m_1 \cdot \omega_1^2 \cdot r_1 \, , \qquad F_2 = m_2 \cdot \omega_2^2 \cdot r_2$$

$$F_1 = m_1 \cdot v_1^2 / r_1 \, , \qquad F_2 = m_2 \cdot v_2^2 / r_2 \, .$$

Condition $v_1 = v_2 = v_0$ allows to write

$$F_1 \sim m_1 / r_1 \, , \qquad F_2 \sim m_2 / r_2 \, .$$

Thus, an equality of the centrifugal forces can be written in the kind

$$m_1 / r_1 = m_2 / r_2$$

or by means of the mass, weight and volume flow correspondingly

$$\dot{m}_1 / r_1 = \dot{m}_2 / r_2 \, ; \qquad \dot{G}_1 / r_1 = \dot{G}_2 / r_2 \, ; \qquad Q_1 / r_1 = Q_2 / r_2 \, . \qquad (3)$$

Just in such correlation is the mass, weight and volume flows and the curvature radii of the initial jet branches at its spreading on the flat plate, oriented under angle within the limits $90^0 > \alpha > 0^0$. This correlation ensures dynamical balance of the immediate division zone of the initial jet into its branches and a constancy of static pressure in the triangle $CDE$ zone.

Characteristic feature of variants of the jet interaction with the plate, considered in this text and presented in fig.2, is that owing to the great size of the plate the initial jet branches adjoin to its surface smoothly and utterly. At that a breadth $ED$ of site of a spreading remains constant at any quantity of angle $\alpha > 0$ between the jet axis and the plate surface. A decrease of this angle leads to a decrease of the jet pressure force onto the plate in accordance with expression (2). The jet pressure centre onto the plate is in the very middle of site $DE$ when $\alpha = 90^0$ and both this site and the jet pressure centre are displaced aside from the jet axis under a decrease of angle $\alpha$, forming an eccentricity $e$. A product of the pressure force, applied in the pressure centre, and the eccentricity of the pressure centre relatively the intersection point of the jet axis with the plate surface forms the force moment for a turning of the plate perpendicularly to the jet axis.

Fig.3$a$ shows a diagram of a spreading of the liquid free jet with its cross-section in the kind of a round on the flat plate $AB$, oriented perpendicularly to the jet axis. In a zone of immediate spreading of the jet the diagram gives the jet longitudinal section by a diametrical flatness, relatively of which a spreading is mirror-symmetrical, although in the given case a spreading is quite axisymmetrical. Observation of the liquid jet spreading in this case allows to ascertain that the initial jet full turn on the plate is restricted by diameter $DE = D_0\sqrt{2}$, i.e. an area of a spreading of the jet on the plate is twice as much of the initial jet cross-section area. A thickness of a fan of a spreading of the jet on a boundary of its full turn, with taking into account of equality of the flow velocity on entrance and exit of the spreading zone and of the continuity condition, is determined by equality of the initial jet cross-section area to a lateral surface area of a cylinder by diameter $D_0\sqrt{2}$ on exit of the spreading zone of the jet

$$\pi \cdot D_0^2 / 4 = h_0 \cdot \pi \cdot D_0 \sqrt{2}$$

and this thickness is

$$h_0 = D_0 / \left(4\sqrt{2}\right) = 0.177 D_0 . \tag{4}$$

A constructing of an internal boundary profile of the spreading jet is quite similar to the fig.2$a$ diagram. An external boundary profile of the spreading jet, with a taking into account of the results of the experiment observation and expression (4), is an evolvent of the first quarter of circumference.

In this case, just as in the case of the flat jet in fig.2$a$, the jet pressure force onto a plate is equal to a product of the initial jet velocity head and of the doubled area of the jet cross-section

$$P_c = 2\pi \frac{D_0^2}{4} \cdot \rho \frac{v_0^2}{2} . \tag{5}$$

Fig.3$b$ shows a diagram of a spreading of the liquid free jet with the round cross-section on the flat plate $AB$, oriented to the jet axis under angle $60^0$. In a zone of the immediate spreading of the jet the diagram gives the jet longitudinal section by diametrical flatness relatively of which the spreading is mirror-symmetrical.

A constructing of an internal boundary profile of the spreading jet is quite similar to a fig.2$b$ diagram. For a constructing of the jet external profile in the diametrical flatness it is necessary:
- draw a horizontal line between the vertical line $EO_1$ and $DO_2$ on a height $h_0$ above line $AB$;
- turn this horizontal line anticlockwise relatively a point, placed in its middle, so that a ratio of a height of its ends above line $AB$ should correspond to a ratio of the parts of the initial jet diameter, formed by the point $C$;
- draw two horizontal lines from the ends of the turned horizontal line in a both sides of it;
- conjugate the initial jet external profile with the above drawn horizontal lines by means of the circumference arches (points $O_3$ and $O_4$ is the curvature centers of these arches).

Such constructing ensures a smooth distribution of the jet weight (volume) flow along a perimeter of the jet spreading zone on the plate: from the largest to the least in the diametrical flatness and the average on each side.

The jet pressure force onto the plate is determined by expression

$$P_c = 2\pi \frac{D_0^2}{4} \cdot \rho \frac{v_0^2}{2} \cdot \sin 60^0 \ . \tag{6}$$

The jet pressure centre onto the plate is displaced from the intersection point of the jet axis with the plate by an eccentricity $e$. A product of the force $P_c$ and the eccentricity $e$ is a force moment, turning the plate perpendicularly to the jet axis.

Analysis of the above stated method of a constructing of the jet external boundary profile in the diametrical flatness shows that such constructing can be realized at angle $\alpha$ within the limits $90^0 > \alpha > 35^0$. When $35^0 > \alpha > 0$ the construction diagram for the liquid jet with a round cross-section transforms into the diagram in fig.2b, c.

The author experiments with the water free jet of a round cross-section show a sufficient change of a character of a spreading of such jet at the angle quantity equal $\sim 35^0$. When $\alpha < 35^0$ the more or less uniform flow along the perimeter of the jet spreading zone gives place to the compact jet with the lateral spreading within the limits of a round with diameter $DE$.

Fig.3c shows a diagram of a spreading of the same jet on the same plate, oriented under angle $30^0$ to the jet axis. As it is above stated the construction diagram of a profile of the jet internal and external boundaries in the diametrical flatness quite corresponds to the fig. 2b, c diagram. The jet pressure force onto the plate is decreased in this case up to quantity

$$P_c = \pi \frac{D_0^2}{2} \rho \frac{v_0^2}{2} \sin 30^0 \ .$$

And at the same time the eccentricity $e$ is sufficiently increased in comparison with the fig.2b diagram.

It is necessary to note that the above adduced expressions (1, 2) determining the pressure $p$ quantity in zone $CDE$ is given without the taking into account of the jet energy expenses for a turn of its branches. Accordingly to the well-known results of the experimental research [13] the maximum dimensionless quantity of the head loss coefficient is 8 % of the jet velocity head when the flat plate is oriented perpendicularly to the jet axis. It is very likely a quantity of the mentioned coefficient is connected with a quantity of the plate orientation angle relatively the free liquid jet axis. This connection can be presented in the kind

$$\zeta(\alpha) = (1 - 0.08 \sin \alpha) \ \text{within the limits } 90^0 > \alpha > 0 \ . \tag{7}$$

Thus, with the taking into account of it the expression (2) must be written in the kind

$$p_c = (1 - 0.08 \sin \alpha) \rho \frac{v_0^2}{2} \sin \alpha \ . \tag{8}$$

Expression (8) is equally correct for both the flat jet and the jet with a round cross-section. The head loss coefficient $\zeta(\alpha)$ allows to take into account the energy expenses on the ventilation zone $CDE$ as well as on formation of vortexes at the circulative motion of the jet branches. With the taking into account of expression (7) a velocity of the flat jet branches on exit out of zone of a spreading, when the flat plate is perpendicularly oriented to the jet axis, will be the same quantity, equal to

$$\bar{v}_{ex} = v_0 \sqrt{1 - 0.08 \sin \alpha} = v_0 \sqrt{1 - 0.08} = 0.96 v_0 \ . \tag{9}$$

In the cases, when the plate is oriented to the jet axis under angle within the limits $90^0 > \alpha > 0$, velocity of the jet branches will be somewhat unequal accordingly to the dependences

$$v_1 = \overline{v}_{ex} + (v_0 - \overline{v}_{ex})\frac{\alpha_1}{90^0}; \qquad v_2 = \overline{v}_{ex} - (v_0 - \overline{v}_{ex})\frac{\alpha_2}{90^0}, \qquad (10, 11)$$

where $v_1$ is the passing branch velocity, $\alpha_1 = \alpha$ ; $\alpha_2 = (180^0 - \alpha)$ is the turn angles of the jet branches to the jet axis.

Conformably to the free liquid jet with a round cross-section, when a spreading on the flat plate is going on more or less uniformly along the perimeter similar to a circumference, the expressions (10, 11) give the greatest and the least quantity of velocity in the diametrical flatness of a spreading and at the same time the expression (9) determines the velocity of the lateral spreading.

The energy loss at the above considered interaction of the free liquid jet with the flat plate is determined in the greatest degree by the so-called form drag. The subsequent current – beyond the bounds of zone $DE$ – is going on under the prevalent influence of a friction.

The above stated method allows also to define the jet bifurcation surface, line. In fig.2$a$ the jet bifurcation surface is a flatness presented in the diagram by its track in the kind of perpendicular dropped from the point $C$ to the plate $AB$. In the given case the jet bifurcation flatness coincides with the middle flatness of the initial flat free jet. When an orientation angle of the plate $AB$ to the flat jet axis is less then $90^0$, the jet bifurcation surface turns into a cylindrical surface, presented in the fig.2$b$, $c$ diagram by its track in the kind of a circumference arch $CF$ with a centre in a point of intersection of a line $O_1O_2$ with the plate $AB$ track. In the case of the free liquid jet with a round cross-section the spreading is going on axial symmetrically as it is shown in fig.3$a$. When the plate $AB$ orientation angle to that jet axis is lesser than $90^0$ the jet bifurcation line is curved into the circumference flat arch with centre in a point of intersection of a line $O_3O_4$ with the plate $AB$ track (fig.3$b$) and in a point of intersection of a line $O_1O_2$ with the plate $AB$ track (fig.3$c$).

Now at last it becomes possible to elucidate a question on the dynamic pressure distribution along the line $DE = 2b$ conformably to the fig. 2$a$. The momentum vector of initial jet is $\rho \cdot Q \cdot v_0 = \rho \cdot A \cdot v_0$, where $Q$ is a volume flow, $A$ is the jet cross-section area. Taking into account that the jet branches turn relatively plate $AB$ within the limits $0 \le \alpha_1 \le \alpha$ and $0 \le \alpha_2 \le (180^0 - \alpha)$ the previous expression can be written in the kind

$$\frac{\rho \cdot Q \cdot v_0}{2A} = \rho\frac{v_0^2}{2}\sin^2\alpha . \qquad (12)$$

Thus the dynamic pressure distribution within the limits of a line $DE$ corresponds to a sine square and one presents by itself the pressure real profile, determined by tangential hydrodynamical forces. At that the jet bifurcation line, coinciding in this case with the jet axis, indicates a place in which the sinusoid has a maximum quantity. This quantity is twice as much the uniformly distributed static pressure, determined by expression (2), i.e.

$$0 \le (p_{ti} = 2p_c\sin^2\alpha_i) \le 2p_c, \qquad (13)$$

although areas of these graphs according to the expressions (2) and (13) are equal. Correspondingly the resultants of hydrodynamical forces in these two cases are also equal, i.e. $P_t = P_c$. In this sense the uniformly distributed static pressure is the simplified equivalent of the real profile, expressing a condition of equilibrium of centrifugal forces, developed by the jet branches, and static pressure as the scalar quantities in the closed zone $CDE$ between the jet branches and the plate $AB$. In contrast to it tangential hydrodynamical forces determine not only resultant of these forces but also character of the pressure profile. These features allow quite simply to construct the pressure real profile as it is shown in fig.2$b$, $c$. At that the jet bifurcation line $CF$ indicates a place of location of the pressure real profile maximum; the resultant force quantity is determined by the pressure real profile area; a gravity centre of the pressure real profile area determines a place of location of the resultant force. Conformably to the fig.3$a$ diagram the pressure real profile is a parabola

$$0 \le \left[ p_{ri} = 2 p_c \left( 1 - r_i^2 / R_1^2 \right) \right] \le 2 p_c \,, \tag{14}$$

where $R_1 = DE/2 = D_0 \sqrt{2}/2$ and $0 \le r_i \le R_1$.

The paraboloid volume is equal to the cylinder volume, constructed on a base of the uniformly distributed static pressure profile. The pressure paraboloid volume determines the resultant force quantity. The jet divergence line $CF$ indicates, as before, a place of location of the pressure real profile maximum, i.e. a height of the paraboloid. The paraboloid gravity centre determines a place of location of the resultant force, i.e. the pressure centre.

Thus, in a range $90^0 > \alpha > 0$ the plate undergoes action of the resultant of hydrodynamical forces and of its moment, as a product of this resultant and eccentricity of its application point relatively centre of site $ED$ of a spreading of the jet, turning the plate normally to the jet axis. In the case $\alpha = 90^0$ the resultant force moment of hydrodynamical forces is absent. At the same time such equilibrium state of the plate is dynamical: in this case the interaction of the jet with the plate is accompanied by autooscillations. At that the plate is cyclically swinged round centre of site $ED$ in both sides (alternately clockwise and anticlockwise) and the weight (volume) flow in the jet branches is cyclically changed accordingly to it. Autooscillations of such kind, arising under action of the unrestricted fluid stream onto a symmetrical form body, were for the first time researched by V. Strouhal (1878).

### Discussion of results

The generalized character of the developed graphic-analytical method of a calculation and geometrical construction of the motion trajectories of the free liquid jet in zone of its interaction with the flat plate allows to apply this method for other combinations of the free jet interaction with the flat plate and other cases.

Fig.4 shows diagrammatically the interaction results of the flat free liquid jet with the flat rectangular plate $AD$, oriented perpendicularly to the jet axis and submerged partially into the jet body; the plate side, submerged into jet, is parallel to the lateral surface of the flat jet. In the fig.4$a$ diagram the plate is submerged into the jet body farther of its symmetry flatness. In the fig.4$b$ diagram the plate is submerged into the jet body before the jet symmetry flatness. In fig.4$c$ the plate slightly touches with the jet. A comparison of these diagrams testifies to that a decrease of the plate submergence depth into the jet body leads to a decrease of the deflection angle of the jet free branch and to an increase of its cross-section and accordingly of its weight (volume) flow; at the same time a decrease of the plate submergence depth leads to a decrease of its area, undergoing the velocity head action, and accordingly to a decrease of the jet pressure force on the plate.

Fig.5 shows diagrammatically the interaction results of the free liquid jet of a round cross-section with the flat rectangular plate $AD$ under the submergence condition of fig.4$a$, $b$ and $c$. The basic difference of the diagrams in fig.5 is that a cross-section of the jet free branch, cut off by the plate, has in its beginning the round segment form.

Fig.6 shows diagrammatically the interaction results of the flat free liquid jet with the flat plate $AD$, oriented under different angles to the jet axis and under different degree of its submergence into the jet body within the limits of the contours $CE$ and $CD$ of its spreading. A comparison of the fig.6$a$, $b$ and $c$ diagrams shows the combined influence of the plate orientation angle to the jet axis and of the submergence degree into the jet body on the jet pressure force upon the plate.

Fig.7$a$ shows a combined diagram of a spreading of the flat free liquid jet on a flat fore part of the rectangular solid body with the different thickness: greater, equal and lesser of the jet thickness.

Fig.7$b$ shows a combined diagram of a spreading of the free liquid jet of a round cross-section on a flat fore part of a cylindrical solid body with diameter greater, equal and lesser of the jet diameter.

Fig.8*a* shows a scheme of interaction of the flat free liquid jet with the flat body fore part in the kind of semicircular cylinder. A feature of this initial scheme is in that the body thickness and accordingly the semicircular cylinder diameter twice as much the jet thickness. In this case a zone of immediate mirror-symmetrical spreading of the jet is restricted by angle $30^0$ in both sides from the jet axis and one is inside of equilateral $OO_1O_2$ triangle. In this zone the jet renders a positive pressure onto the body fore part. At further flow within range $30^0 - 40^0$ the centrifugal forces of the jet branches render a negative pressure onto the body fore part. A breadth of $DE$ zone is equal to the jet thickness - $DE = 2b$, therefore a pressure force quantity of the jet onto this part of the cylindrical surface is approximately half a pressure force when the jet spreads on a flat plate. In this part the obtained here results quite correspond to the results of O. Flachsbart experiments [14], conducted by means of interaction of the unrestricted fluid stream with a body in the kind of cylinder.

Fig.8*b* shows a scheme of interaction of the round free liquid jet with the hemispherical fore part of a cylindrical body. A feature of this initial scheme is in that the body cross-section area and accordingly the middle area of the body hemispherical fore part twice as much the jet cross-section area. In this case a zone of the immediate axisymmetrical spreading of the jet around the hemisphere, the same way as in the fig.9*a* scheme, is restricted by angle $30^0$ from the jet axis. Side by side with it in contrast to a scheme in fig.9*a* an action zone of the jet positive pressure onto the hemisphere, determined in this case additionally by the jet external profile, is restricted by angle within the range $40^0 - 45^0$ from the jet axis. Other distinction from a scheme in fig.9*a* is in that an area of $DE$ zone is half the initial jet cross-section area; therefore a pressure force quantity of the jet onto this sphere part is one quarter of a pressure force when the jet spreads on a flat plate. In this part the obtained here results do not contradict Flachsbart experiments [15], conducted by means of interaction of the unrestricted fluid stream with a body in the kind of sphere. Absence of contradictions between results, obtained in the given article on the one hand, and results, obtained by Flachsbart on the other hand, testifies to that an interaction of the free liquid jet with the body fore part is more simple in comparison with an interaction of the unrestricted fluid stream with the body as whole. Hence it follows: interaction of the unrestricted fluid stream with a streamlined body is accompanied by origin of the jet stream – before the body fore part and of the jet layer – along the body fore part. This jet layer possesses the velocity head quantity equal to the jet stream in its velocity head. This jet layer interacts with the body surface and with the unrestricted fluid stream and just this jet layer determines a flow field around the body. Just such flow field structure is inherent to the real fluid motion.

The generalized character of the developed method allows also its applying for a solving of the problems, bound with interaction of the jets between themselves and others.

Fig.9 demonstrates the crown structure of the liquid fountain with the jet of a round cross-section. This flow field allows to understand why the comparatively not great and sufficiently light ball can be held out on the fountain crown [16, 17]. At the same time the not great declination of the fountain jet from vertical line demonstrates a spreading of its crown similar to the fig.3*b* diagram. This circumstance testifies to that the water fountain compact jet experiences the brake blow on one's reaching of maximum height, although the solid barrier is absent on its path. The other feature of the fountain compact jet, moved through the rest atmosphere, is in increase of its diameter in compliance with its height and in the corresponding decrease of its height in comparison with the theoretical quantity, determined by the so-called Torricelli formula. A cause of this second feature is bound with friction of the moving jet against the rest atmosphere. Let us consider the following simplified example. The fountain compact jet flows out of a round mouthpiece 0.1m diameter with $v_0 = 10$ m/s initial average velocity and one moves vertically upwards.

In the absence of friction between the liquid jet and atmosphere the fountain height is determined by Torricelli formula

$$h_0 = \frac{1}{g}\frac{v_0^2}{2} = \frac{1}{9.81}\frac{10^2}{2} = 5.1 \, \text{m.}$$

The fountain jet has a cylindrical form in this case.

In the presence of friction between the jet and the immovable atmosphere a decreasing of the fountain jet velocity up to its top is determined by expression

$$\Delta v = \sqrt{2g\Delta p/\gamma_l} = v\sqrt{\lambda\frac{h_0}{d_0}v_0^2\frac{\gamma_a}{\gamma_l}} = \sqrt{0.02\frac{5.1}{0.1}10^2\frac{1.23}{10^3}} = 0.354 \, \text{m/s,}$$

where $\Delta p = \lambda\frac{h_0}{d_0}\frac{\gamma_a}{g}\frac{v_0^2}{2}$ is the jet head loss because of its friction against air by Darcy –

– Weissbach; $\gamma_l$ is the water density, $\gamma_a$ is the air density, $\lambda$ – coefficient of hydraulic friction of atmospheric air against the liquid jet, determined by Reynolds number

$$\text{Re}_a = d_0\frac{\gamma_a}{g}\frac{v_0}{\mu_a} = 0.1\frac{1.23}{9.81}\frac{10}{1.9}10^6 = 66000$$

(here $\mu_a = 1.9 \cdot 10^{-6} kgf \cdot s/m^2$ is the air dynamic viscosity coefficient), and its quantity, in accordance with Colebrook's diagram, makes 0.02.

In the result the fountain jet velocity at its top is decreased up to quantity

$$v_h = v_0 - \Delta v = 10 - 0.354 = 9.646 \, \text{m/s.}$$

Accordingly the fountain jet diameter is increased up to

$$d_h = d_0\sqrt{\frac{v_0}{v_h}} = 0.1\sqrt{\frac{10}{9.646}} \approx 0.104 \, \text{m.}$$

Thus, in the presence of a friction the fountain compact jet has a form similar to the truncated cone. According to a continuity condition the fountain jet volume in both cases considered in the given example is of the same quantity, i.e.

$$h_0\pi\frac{d_0^2}{4} = h_{tc}\frac{\pi}{12}\left(d_0^2 + d_h^2 + d_0 d_h\right);$$

hence the fountain jet full height with the taking into account of a friction is

$$h_{tc} = h_0\frac{3 \cdot d_0^2}{d_0^2 + d_h^2 + d_0 d_h} = 5.1\frac{3 \cdot 0.1^2}{0.1^2 + 0.104^2 + 0.1 \cdot 0.104} = 4.904 \, \text{m}$$

and one includes its spreading crown by height equal approximately $d_h$.

With a point of view of energy conservation the given example allows to conclude: a friction transforms a part of the longitudinal velocity head of the fountain liquid jet in the rest atmosphere into its transversal velocity head in proportion to the friction intensity; at that the total head of such jet is a geometrical sum of its longitudinal and transversal heads as the vector quantities. In the result the Total Head Vector (THV) of the jet is deflected from the jet axis, and this Vector determines an inclination of the jet conical surface generatrix under action of the jet friction against atmosphere. In the absence of friction the Total Head Vector coincides with the jet axis and the jet remains to be cylindrical. In both these cases static pressure in the jet remains equal to a pressure of the surrounding atmosphere.

Considering the fountain free jet one cannot but notice a strange character of its motion. On the one hand, the fountain height is determined by Torricelli formula that is just for a free fall. On the other hand, Torricelli's modified formula $v_i = \sqrt{2gh_0\left(1 - h_i/h_0\right)}$, adduced in article [12], and connecting the running height of a body fall with its running velocity, testifies to that the running velocity of the body, thrown vertically upwards, is changed according to a quadratic law, and this velocity will be equal to a zero in upper point. The mentioned formulas are also just for the free-falling liquid jet according to Torricelli experiments. However, for the fountain jet, moving vertically upwards, the mentioned connection of the running velocity with its height contradicts the continuity condition as according to the free fall law the fountain jet velocity and its weight (volume) flow in upper points must be equal to a zero. At the same time it is apparently that the fountain jet keeps its cylindrical form and, consequently, a constancy of the running velocity along the height without a taking into account of a friction. For a solving of the mentioned contradiction it is necessary to elucidate the dynamical state of the fountain jet. In his previous work [3] author the given article has supposed a fluid flow through the flow element in the kind of a pipe as an autooscillating process. Conformably to the fountain jet such approach allows in one turn to suppose the fountain jet structural analogue, composed from the solid bodies:
- a bar, mounted vertically on a bearing;
- the bearing, provoking the longitudinal harmonic oscillations in this bar;
- an additional mass, mounted on the upper end of the bar and elastically connected with it.

Such system needs in energy of the forced oscillations of the bearing for its functioning. In contrast to it an interaction of the fountain jet with a fluid stream in the flow element (pipe) excites in itself the longitudinal elastic oscillations both in the jet column and in the pipe stream. In this case the fountain jet column as the bar presents by itself a waveguide conducting the longitudinal elastic waves to its top and exciting oscillations of the liquid additional mass. Interaction of the fountain jet column with its liquid additional mass on its top leads to a forming of a spreading crown as it is shown in the fig.9 diagram. In a result of it a height of the fountain jet corresponds to the free fall law and at the same time velocity and a cross-section area of the fountain jet column do not correspond to the quadratic law of a free fall and ones keep a constant quantity in absence of its friction against atmosphere.

The considered feature of the fountain jet, side by side with the examples of the laminar flow instability [7] and the necking in a bar of plastic material under its tension [2], testify to that the presence of the freedom additional degrees of the fluid motion in comparison with a solid body in the kind of internal mobility does not abolish the freedom degrees of the body motion and combines in its motion the freedom additional degrees with the freedom initial degrees, what are inherent in solid body. In other words: mechanical activity of a liquid is much more in comparison with solid body; and gas, in addition to the mechanical activity of a liquid, combines in itself the thermodynamical activity; in ones turn, a plasma, in addition to the gas mechanical and thermodynamical activity, combines in itself electrical activity. This thesis is one of the bases, ensuring an effectiveness of the Physical Ensemble method for solution of problems of technical physics and the fluid motion in particular.

Fig.10. demonstrates an interaction diagram of two flat free liquid jets with the same thickness and the weight (volume) flow, moving coaxially in the opposite directions.

Fig.11. shows an interaction diagram of two flat free liquid jets, moving coaxially in the opposite direction; the left jet has a third of thickness and the weight (volume) flow of the contrary jet.

Fig.12. demonstrates an interaction diagram of two flat free liquid jets with the same thickness and weight (volume) flow, moving under angle $120^0$ against each other; on the left it is conditionally showed after-part of a streamlined flat body.

Fig.13. shows scheme of interaction of two flat free liquid jets with a ratio of its thickness equal 1:3,

moving against each other along the plate AB under the weightlessness conditions; in this case the external contour of a flow coincides with the external contour of a flow, shown in fig.2b, although the flow direction is reverse.

Fig.14. shows an interaction diagram of two free liquid jets with a round cross-section and with the same weight (volume) flow, moving under angle $120^0$ against each other.

Fig.15. shows an interaction diagram of two free liquid jets of a round cross-section, moving coaxially against each other and forming a conical liquid shroud.

Fig.16. shows confluence of a conical liquid shroud into two round jets, moving in opposite directions; on the left it is conditionally showed after-part of a streamlined axisymmetrical body.

Fig.17 demonstrates Rankine's flat hydrodynamical body structure, formed by interaction of the flat jet, flowing out of a flat channel, with the contrary unrestricted liquid stream, constructed by means of above stated graphic-analytical method. This structure corresponds to fig.2 of an Album [18].

Fig.18 demonstrates Rankine's axisymmetrical hydrodynamical body structure, formed by interaction of the liquid round jet, flowing out of a pipe, with the contrary unrestricted liquid stream. This structure is identical to fig.22 of an Album [18] and one elucidates the erosive action dynamics of the fluid jets, for example such as hydraulic monitor.

H. N. Abramovitch in his book [19] has adduced the results of experimental researches of a bending of axis of the air jets with the flat and round cross-section, blown into the half-restricted air stream across to its motion, and the calculation methods on a base of these experiments with use of a mean velocity of the jets and the stream in these calculations. According to the calculations, the equation of the bent axis of the flat jet has the kind

$$y = \left(2/k\right) \cdot \left( \pm \sqrt{kx + \cot^2 \alpha} - \cot \alpha \right),$$

where $k = \left(w/v_0\right)^2 \cdot c_n / \left(\delta_0 \cdot \sin^2 \alpha\right)$ and $w, v_0$ are the air stream velocity and initial velocity of the air jet, respectively; $c_n$ is a dimensionless coefficient of the jet hydraulic resistance; $\delta_0$ is a half-breadth or radius of the jet cross-section; $\alpha$ is the inclination angle of the air jet to the motion direction of the air stream.

When $\alpha = 90^0$ and $c_n = 1$ the afore-cited well-known equation can be reduced to the kind

$$y = \sqrt{2}\left(v_0/w\right)\sqrt{2 \cdot \delta_0 \cdot x},$$

and when $\left(v_0/w\right) = \sqrt{2}/2$ this equation takes the kind of natural equation of the circumference evolvent, formally analogous to Torricelli formula,

$$y = \sqrt{2 \cdot \delta_0 \cdot x},$$

which one is a base of a constructing of evolventic parabola according to the diagram in fig.1 of the given article. This evolventic parabola divides all quadratic parabolas into two multitudes:

- epievolventic parabolas $\left(O_1 P_{eev}\right)$ when $\left(v_0/\ w\right) > \left(\sqrt{2}/2\right)$;
- hypoevolventic parabolas $\left(O_1 P_{hev}\right)$ when $\left(v_0/w\right) < \left(\sqrt{2}/2\right)$.

Analysis of the results of experimental researches, adduced in figs.12.25 and 12.27 of the above-mentioned book [19], has corroborated an existence of two such multitudes of parabolas and an accordance of the motion trajectories of real jets to this multitude of parabolas.

Thus, solution of a problem on the interaction mechanics of the free liquid jet with the flat plate and its application examples to the interaction of the jets between themselves and others, adduced in the given article, testify to that now – more then 200 years after Du Buat's experimental researches – the united physically substantial and mathematically sufficiently strict method for solution of the given sufficiently wide range of an engineering problems is at long last created.

**Final remarks**

Simplicity of the method, stated in the given article, quite corresponds to Newton expression: *... for Nature is pleased with simplicity, and affects not the pomp of superfluous causes*. Today it can be added: hydrodynamics is the classical mechanics field, verifying the physical strength of scientist, scholar.

**Acknowledgements**

Author wants to express his profound respect to the outstanding precursors, supposing that the modern results, filling the mechanics separate gaps, are stipulated by the natural development of the great heritage. Author wants also to express his deep gratitude to his son Alexei and granddaughter Xenia for a carrying out of this article graphical part and his daughter Catherine for her caring for her father.

---

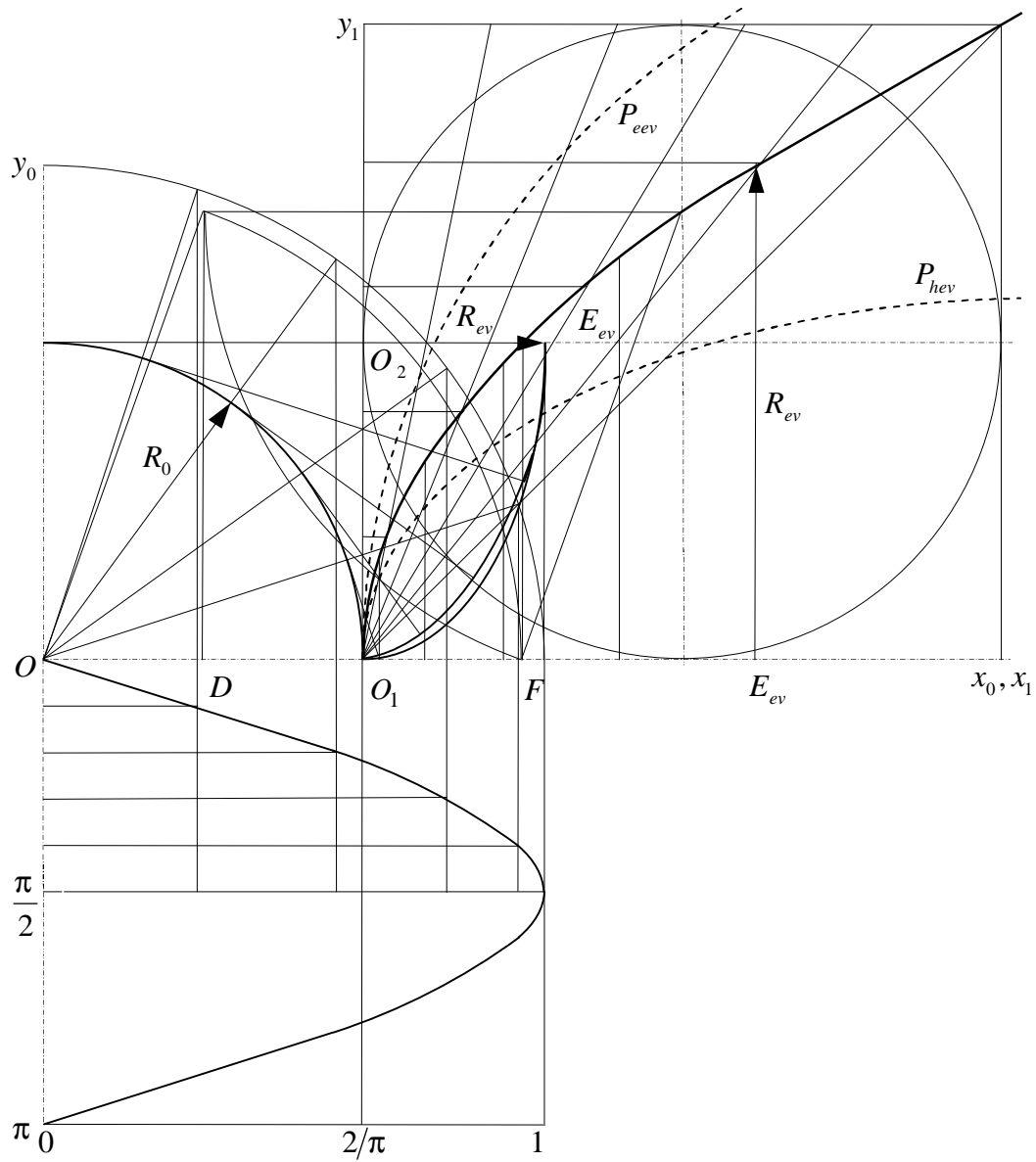

Fig.1. Family of curves – evolvent, parabola, sinusoid, ellipse – constructed on a base of a circumference quarter and used in the given article:

$R_0$ - radius of the initial circumference quarter; $O_1E_{ev}$ - evolvent of the initial circumference quarter; $R_{ev} = R_0\pi/2$ - the maximum radius of the first quarter of the evolvent; $F$ - focus of the evolventic parabola $O_1P_{ev}$, constructed by means of a straightening of the evolvent $O_1E_{ev}$ along the abscissa axis and of a raising of the evolvent radii perpendicularly to it; the evolvent $O_1E_{ev}$ length is $R_{ev}^2/(2R_0) = R_0\pi^2/8 \approx 1.234R_0$; vertical line via the point $D$ is a directrix of the parabola $O_1P_{ev}$: $OD = DO_1 = O_1F = R_0/2$; curve $O_1P_{eev}$ is epievolventic parabola, curve $O_1P_{hev}$ is hypoevolventic parabola; if $R_{ev} = 1$ as an amplitude of harmonic oscillations, then $OO_1 = R_0/R_{ev} = 2/\pi$ is the mean quantity of the amplitude of harmonic oscillations during one-half period; $O_1E_{el}$ - the ellipse quarter as the circumference quarter projection under angle ~55.2$^0$ (point $E_{el}$ coincides with point $E_{ev}$)

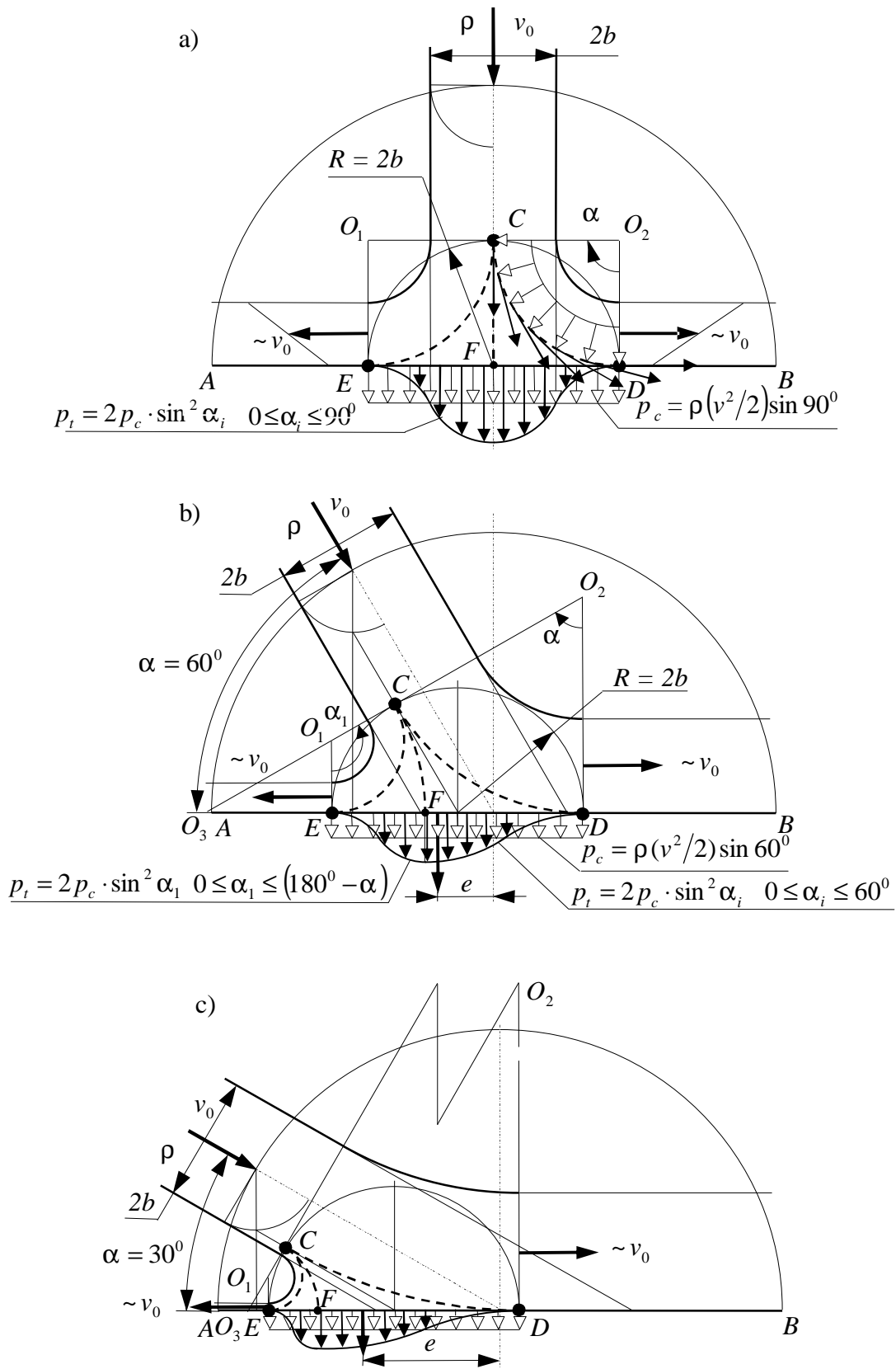

Fig.2. Diagrams of a spreading of the flat free liquid jet on the flat plate *AB*, oriented to the jet axis:
a) perpendicularly; b) under angle $60^0$; c) under angle $30^0$;
$p_c$ - centrifugal pressure;  $p_t$ - tangential pressure

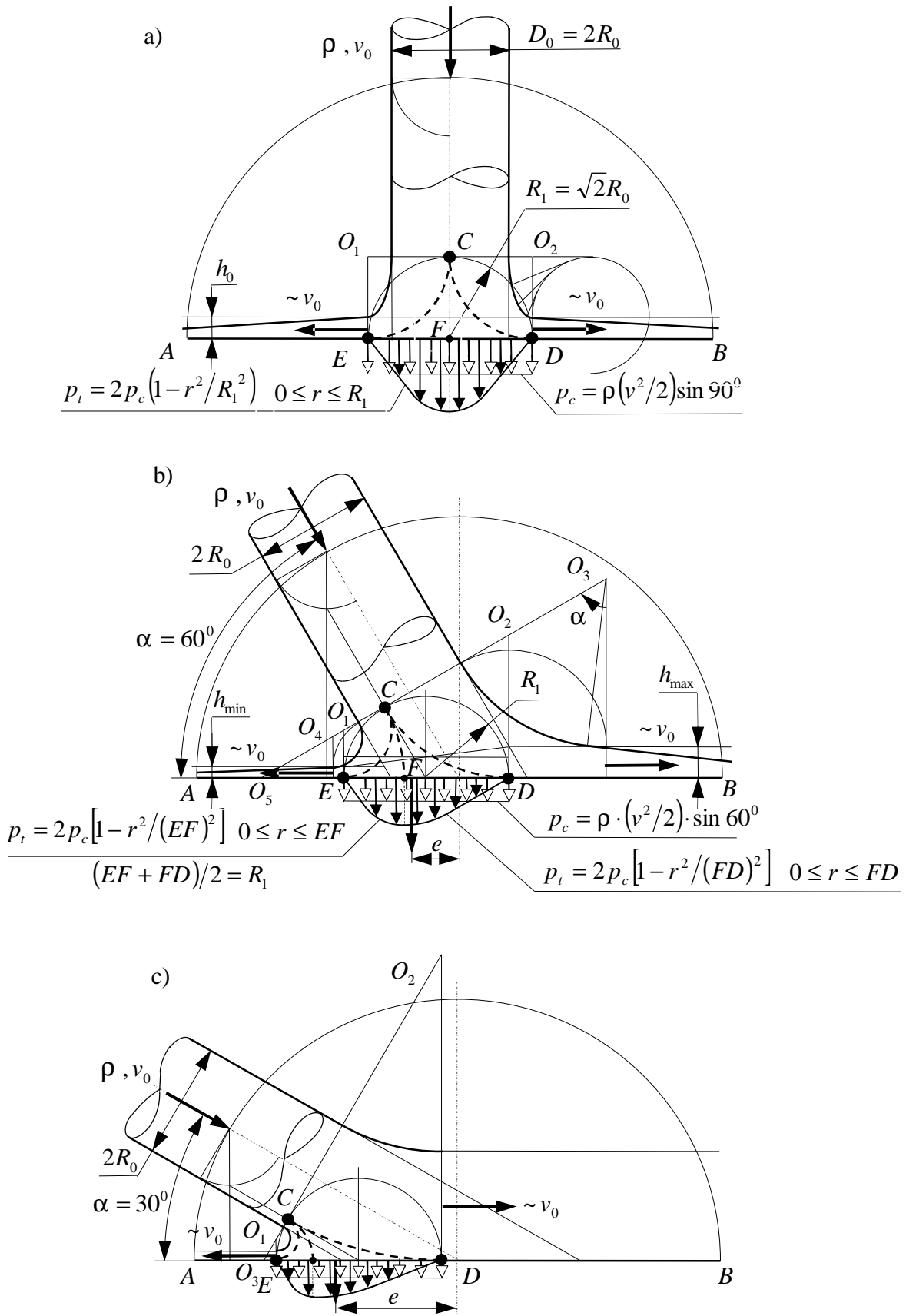

Fig.3. Diagrams of a spreading of the free liquid jet with a round cross-section on the flat plate *AB*, oriented to the jet axis: a) perpendicularly; b) under angle $60^0$; c) under angle $30^0$

Fig.4. Diagram of a spreading of the flat free liquid jet on the flat rectangular plate, oriented perpendicularly to the jet axis and submerged partially into the jet body

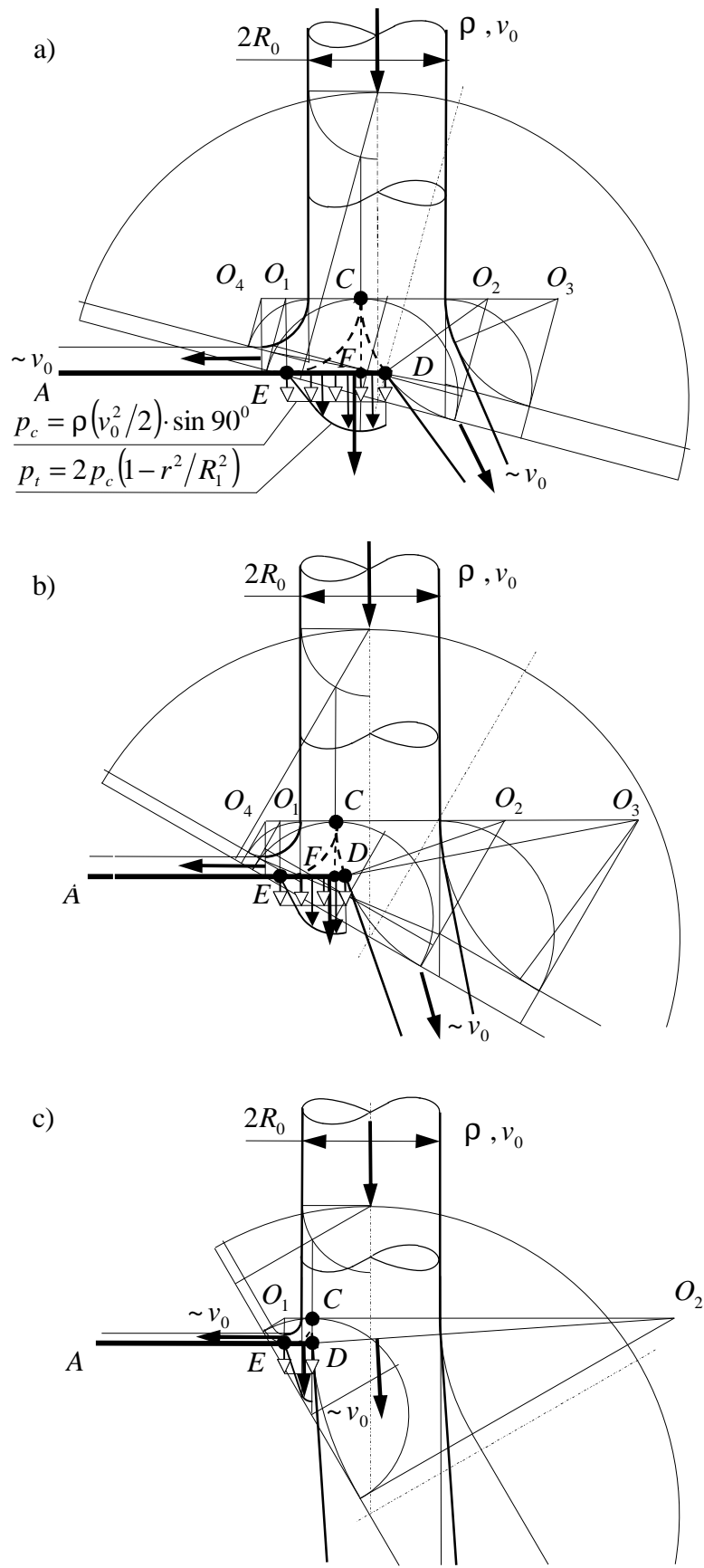

Fig.5. Diagrams of a spreading of the free liquid jet with a round cross-section on the flat plate, oriented perpendicularly to the jet axis and submerged partially into the jet body

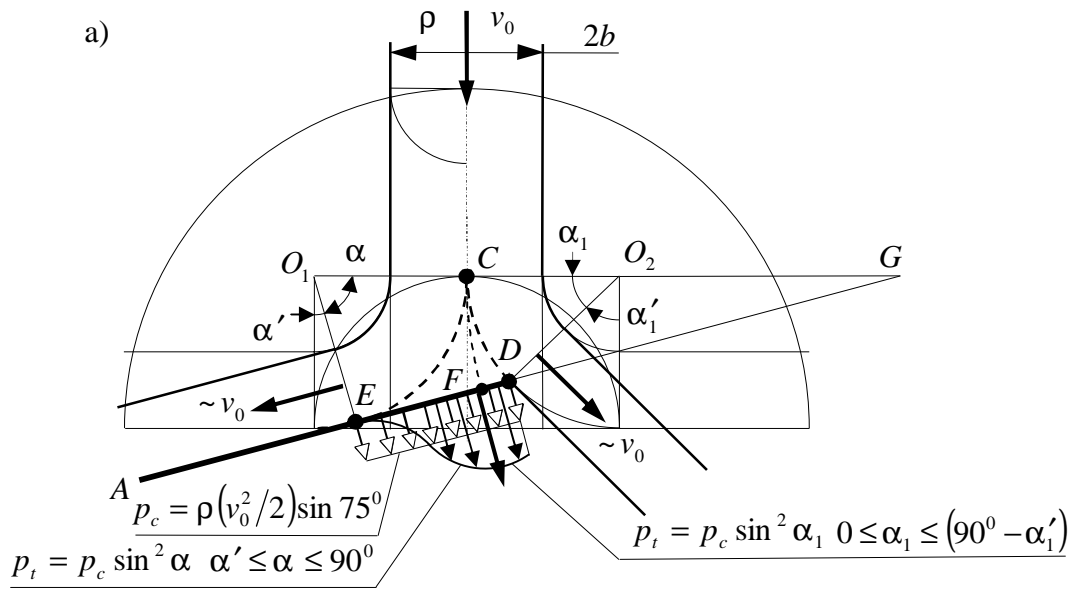

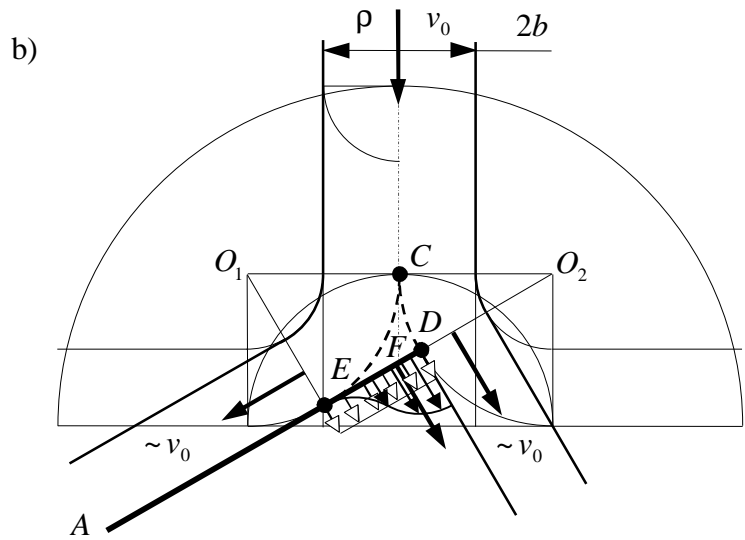

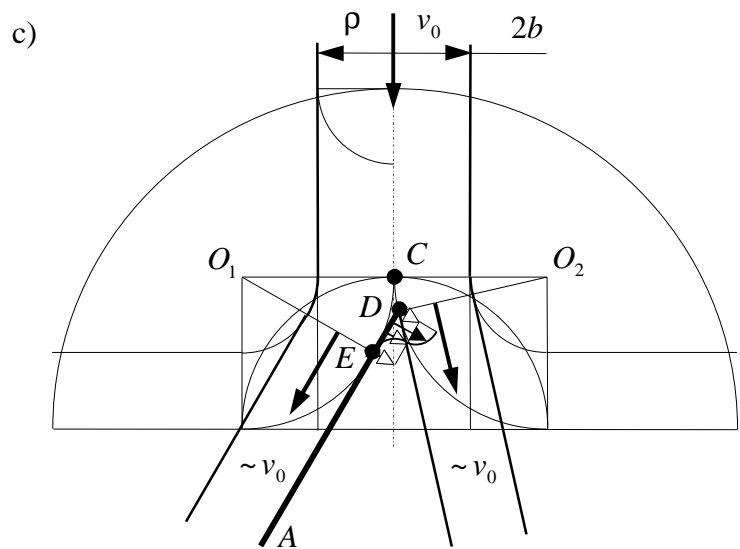

Fig.6. Diagrams of a spreading of the flat free liquid jet on the flat plate AD, partially submerged into the jet body and oriented to the jet axis under angle: a) $75^0$; b) $60^0$; c) $30^0$

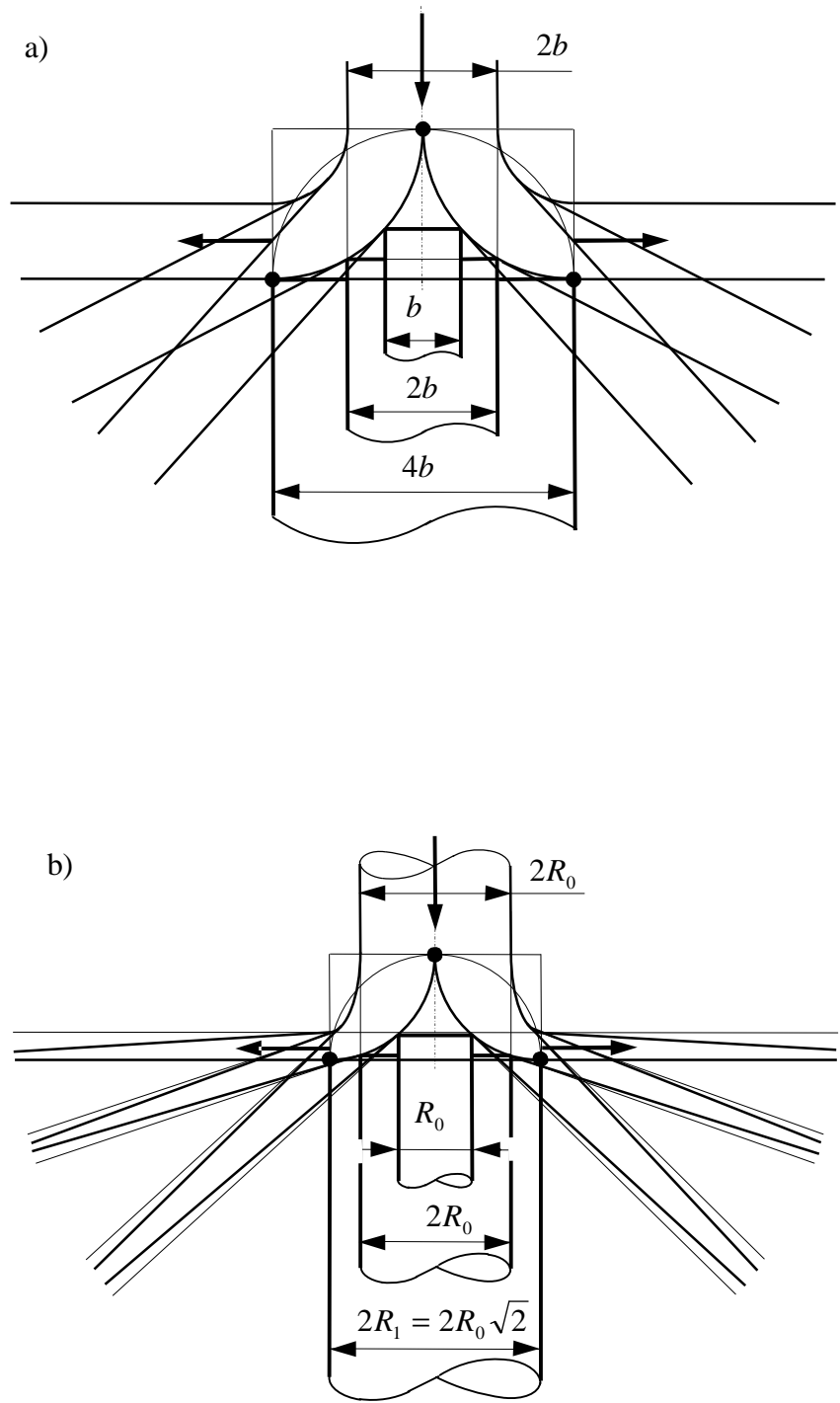

Fig.7.The combined diagrams of a spreading of the flat free liquid jet (a) on the rectangular solid body and of the liquid free jet of a round cross-section (b) on the cylindrical solid body, placed coaxially to the jet

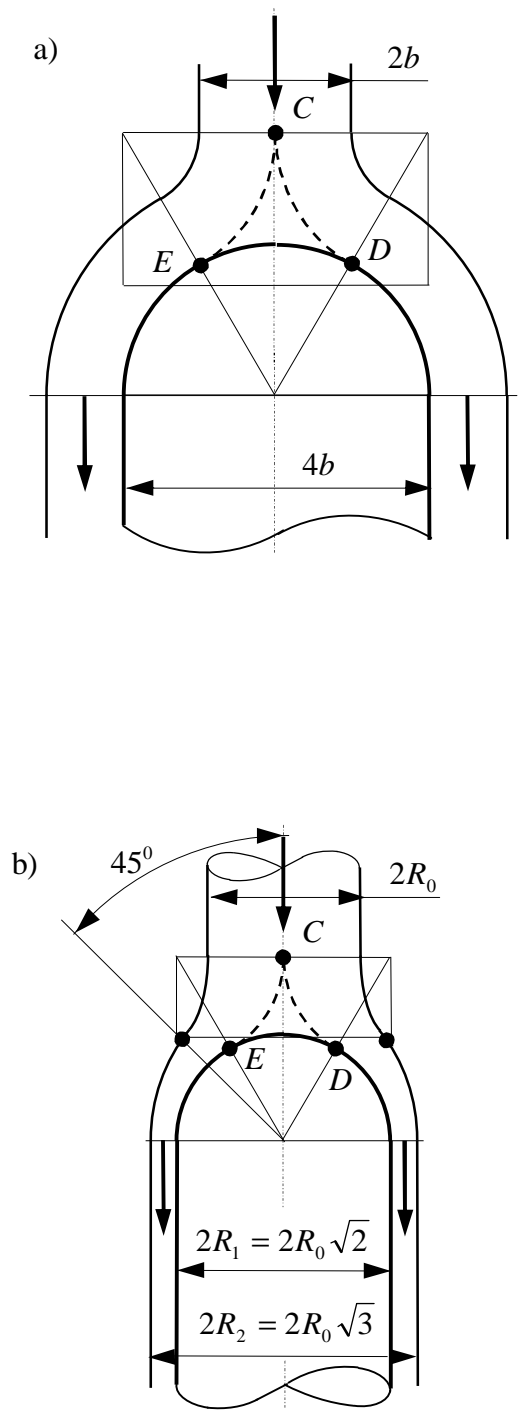

Fig.8. Schemes of interaction of the flat free liquid jet with the flat body fore part in the kind of semicircular cylinder (a) and of interaction of the round free liquid jet with the hemispherical fore part of a cylindrical body (b)

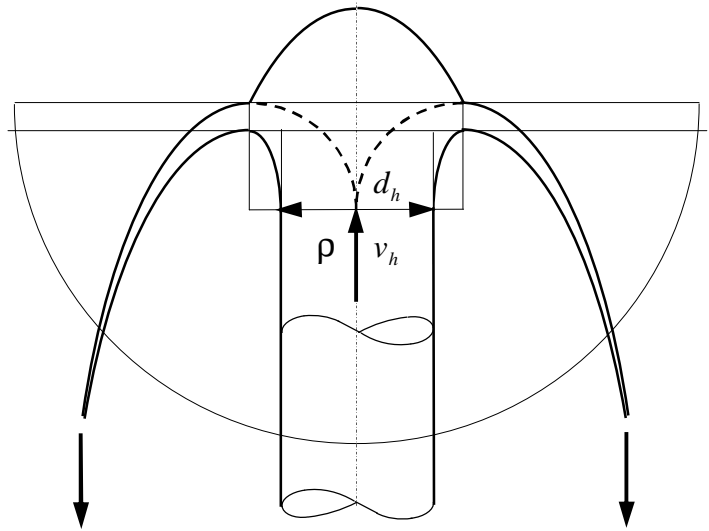

Fig.9. A crown structure of the water fountain jet, constructed by means of a circumference quarter, of a circumference quarter evolvent and of a quadratic parabola

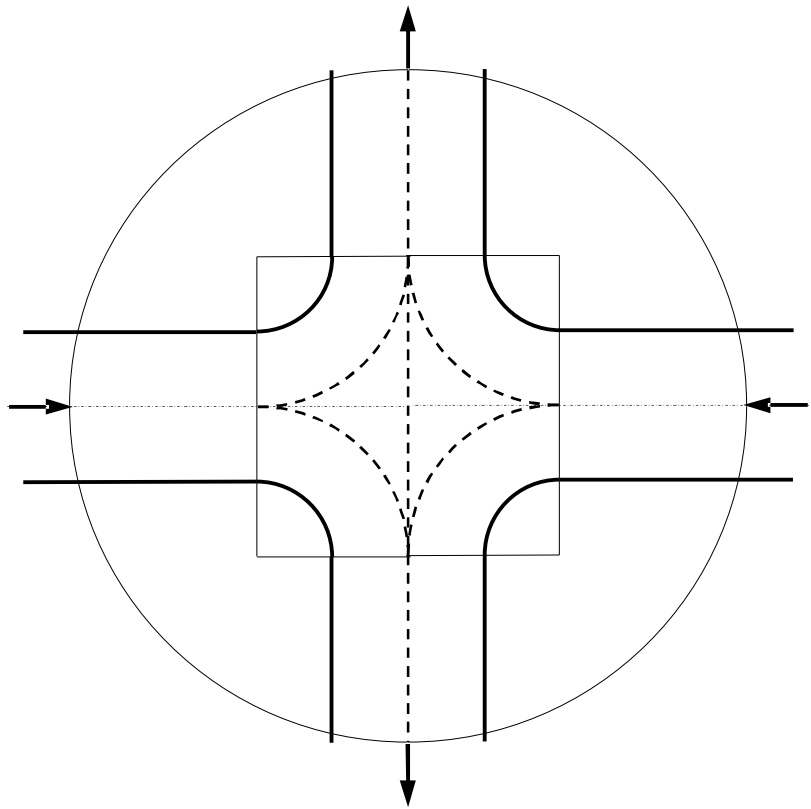

Fig.10. Interaction of two flat free liquid jets with the same thickness and the weight (volume) flow, moving coaxially in the opposite directions

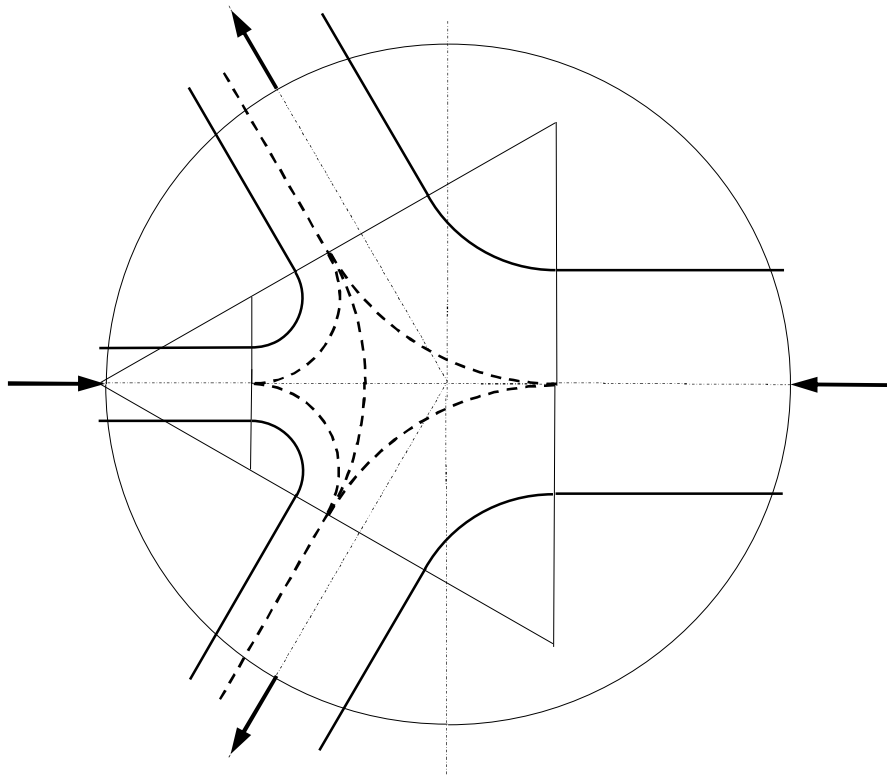

Fig.11. Interaction of two flat free liquid jets, moving coaxially in the opposite direction; the left jet has a third of thickness and the weight (volume) flow of the contrary jet

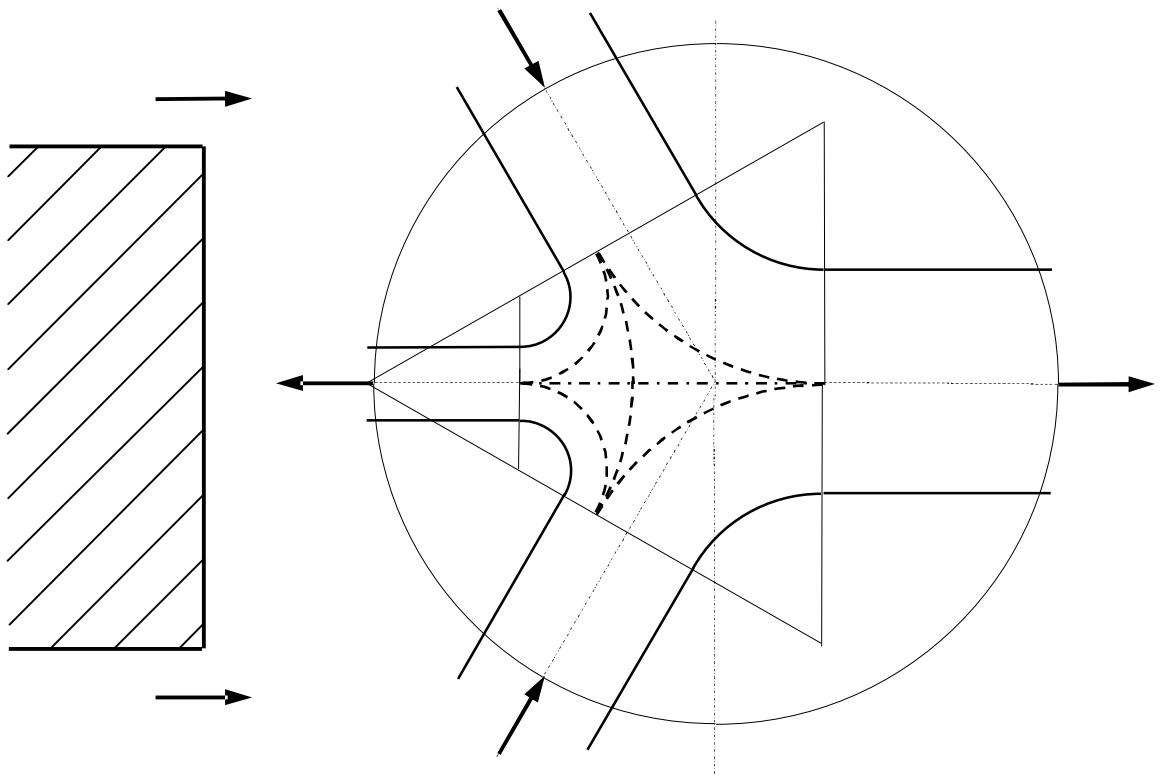

Fig.12. Interaction of two flat free liquid jets with the same thickness and weight (volume) flow, moving under angle $120^0$ against each other; on the left it is conditionally showed after-part of a streamlined flat body

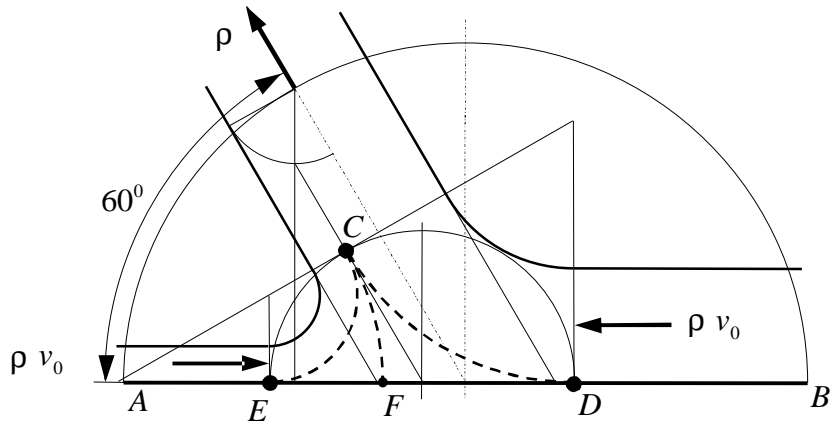

Fig.13. Scheme of interaction of two flat free liquid jet with a ratio of its thickness equal 1: 3, moving against each other along the plate AB under the weightlessness conditions; in this case the external contour of a flow coincides with the external contour of a flow, shown in fig.2*b,* although the flow direction is reverse

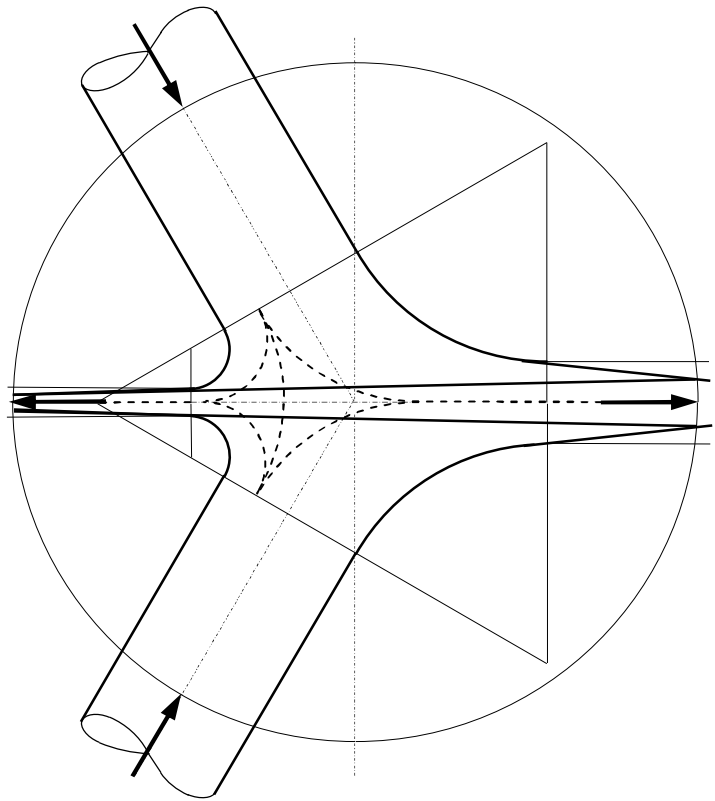

Fig.14. Interaction of two free liquid jets with a round cross-section and with the same weight (volume) flow, moving under angle $120^0$ against each other

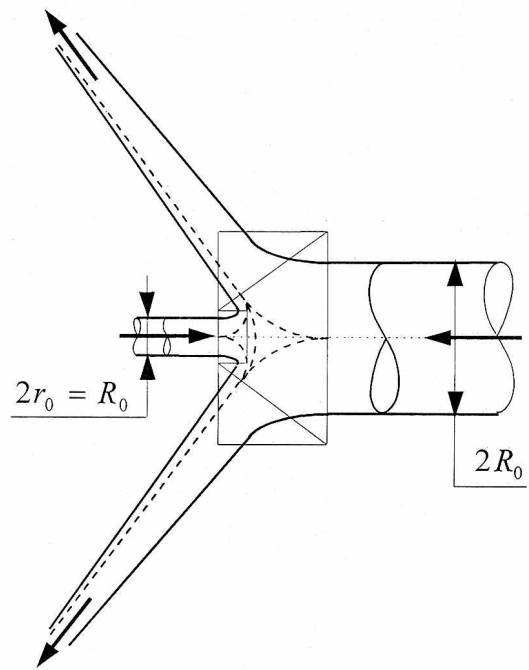

Fig.15. Interaction of two free liquid jets of a round cross-section, moving coaxially against each other and forming a conical liquid shroud

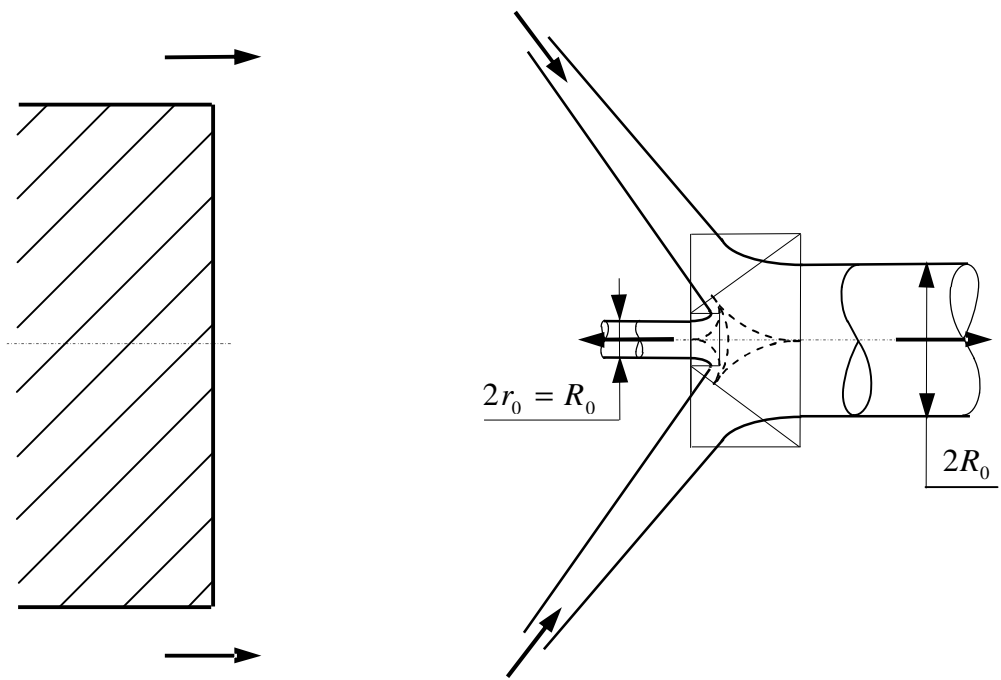

Fig.16. Confluence of conical shroud into two round jets, moving in opposite directions; on the left it is conditionally showed after-part of a streamlined axisymmetrical body

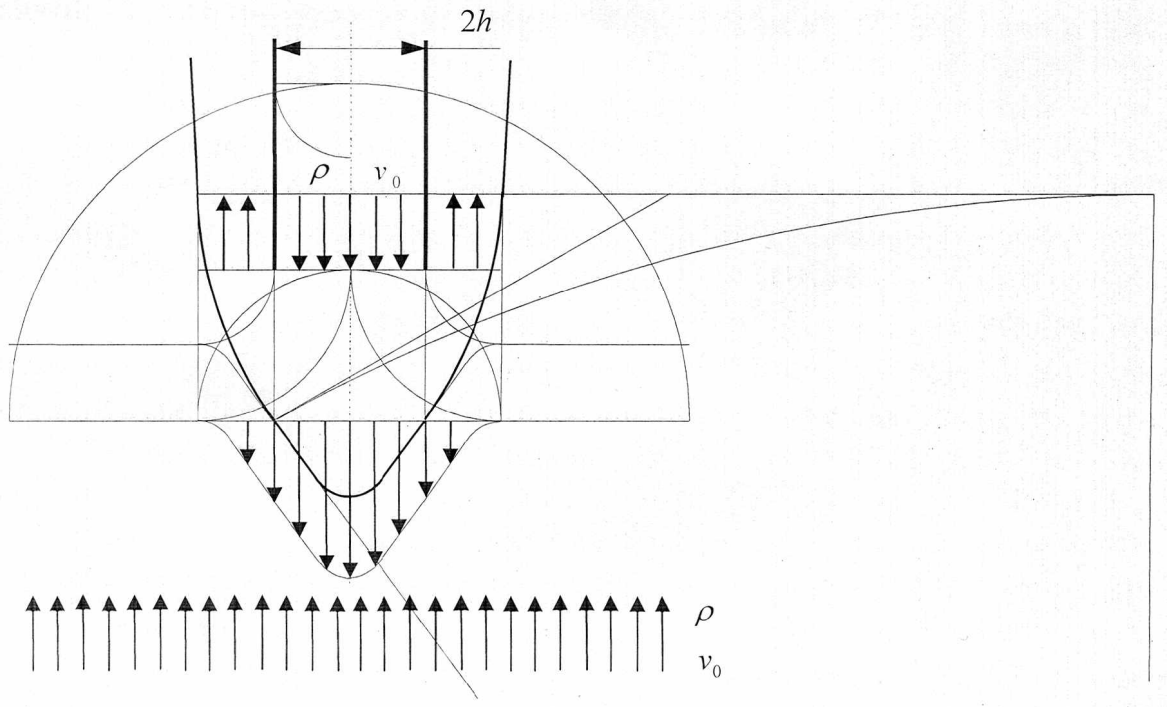

Fig.17. The profile structure of Rankine's flat hydrodynamical body: an interaction surface of the flat liquid jet with the unrestricted liquid stream, directed oppositely to the jet, is formed by the sinusoidal straight cylindrical surface, then by the evolventic straight cylinder and further by the inclined flat surface

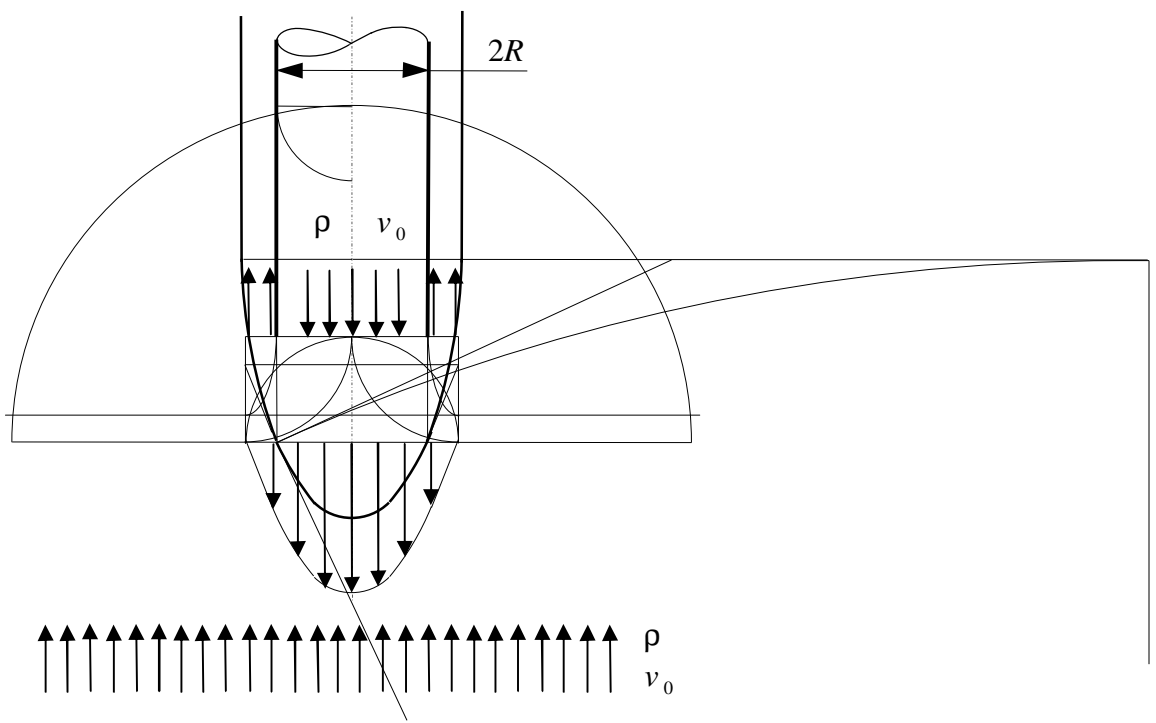

Fig.18. The profile structure of Rankine's axisymmetrical hydrodynamical body: an interaction surface of the round liquid jet, flowing out of socket, with the unrestricted liquid stream, directed oppositely to the jet, is formed by the rotation paraboloid, then by the rotation evolventoid and further by a surface closed to a cone